**ANKARA UNIVERSITESI FEN FAKULTESI FIZIK BOLUMU**

# Why the Four SM Families

S. Sultansoy

Deutsches Elektronen-Synchrotron DESY, Hamburg, Germany
Department of Physics, Faculty of Sciences, Ankara University, Turkey
Institute of Physics, Academy of Sciences, Baku, Azerbaijan

**Abstract**

The flavor democracy favors the existence of the nearly degenerate fourth SM family, whereas the fifth SM family is disfavored both by the mass phenomenology and precision tests of the Standard Model. The multi-hundreds GeV fourth family quarks will be copiously produced at the LHC. However, the first indication of the fourth family may come from the Higgs search at the upgraded Tevatron.

---

Electronic addresses: sultanov@mail.desy.de
sultan@science.an                kara.edu.tr

## 1. Introduction

Today, the mass and mixing patterns of the fundamental fermions are the most mysterious aspects of the particle physics. Even the number of fermion generations does not fixed by the Standard Model (SM). In this sense, SM may be treated as an effective theory of fundamental interactions rather than fundamental particles. The statement of the Flavor Democracy (or, in other words, the Democratic Mass Matrix approach) [1], which is quite natural in the SM framework, may be considered as the interesting step in true direction. It is intriguing, that Flavor Democracy favors the existence of the fourth SM family [2, 3]. Moreover, Democratic Mass Matrix approach provide, in principle the possibility to obtain the small masses [4, 5] for the first three neutrino species without see-saw mechanism. The fourth family quarks, if exist, will be copiously produced at the LHC [6]. Then, the fourth family leads to an essential increase [7, 8] of the Higgs boson production cross section via gluon fusion at hadron colliders and this effect may be observed soon at the Tevatron.

In this letter we consider the present status of the four family SM and the future search for the fourth family fermions. The Flavor Democracy hypothesis is presented in Section 2, where we also give some predictions for the fourth family phenomenology. Arguments against the fifth SM family are listed in Section 3. In Section 4 possible manifestations of the fourth family fermions at future colliders are considered. Possible alternative scenarios such as additional Higgs doublets (the MSSM case) and exotic new fermions ($E_6$ phenomenology) are briefly discussed in Section 5. Finally, in Section 6 we give some concluding remarks.

## 2. Flavor Democracy and the Standard Model

It is useful to consider three different bases:

- Standard Model basis $\{f^0\}$,
- Mass basis $\{f^m\}$ and
- Weak basis $\{f^w\}$.

According to the three family SM, before the spontaneous symmetry breaking quarks are grouped into following $SU(2) \times U(1)$ multiplets:

$$\begin{pmatrix} u_L^0 \\ d_L^0 \end{pmatrix}, u_R^0, d_R^0 \; ; \; \begin{pmatrix} c_L^0 \\ s_L^0 \end{pmatrix}, c_R^0, d_R^0 \; ; \; \begin{pmatrix} t_L^0 \\ b_{Ll}^0 \end{pmatrix}, t_R^0, b_R^0 \,. \quad (1)$$

In one family case all bases are equal and for example, d-quark mass is obtained due to Yukawa interaction

$$L_Y^{(d)} = a_d \left( \overline{u}_L \; \overline{d}_L \right) \begin{pmatrix} \varphi^+ \\ \varphi^0 \end{pmatrix} d_R + h.c. \Rightarrow L_m^{(d)} = m_d \overline{d} d \,, \quad (2)$$

where $m_d = a_d \eta$, $\eta = <\phi^0> \cong 249$ GeV. In the same manner $m_u = a_u \eta$, $m_e = a_e \eta$ and $m_{ve} = a_{ve} \eta$ (if neutrino is Dirac particle). In $n$ family case

$$L_Y^{(d)} = \sum_{i,j=1}^{n} a_{ij}^d \begin{pmatrix} \overline{u}_{Li}^0 & \overline{d}_{Li}^0 \end{pmatrix} \begin{pmatrix} \varphi^+ \\ \varphi^0 \end{pmatrix} d_{Rj}^0 + h.c. = \sum_{i,j=1}^{n} m_{ij}^d \overline{d}_i^0 d_j^0, \quad m_{ij}^d = a_{ij}^d \eta, \quad (3)$$

where $d_1^0$ denotes $d^0$, $d_2^0$ denotes $s^0$ etc. The diagonalization of mass matrix of each type of fermions, or in other words transition from SM basis to mass basis, is performed by well-known bi-unitary transformation. Then, the transition from mass basis to weak basis result in CKM matrix

$$U_{CKM} = (U_L^u)^+ U_L^d,$$

which contains 3(6) observable mixing angles and 1(3) observable CP-violating phases in the case of three (four) SM families.

Before the spontaneous symmetry breaking all quarks are massless and there are no differences between $d^0$, $s^0$ and $b^0$. In other words fermions with the same quantum numbers are indistinguishable. This leads us to the *first assumption* [1], namely, Yukawa couplings are equal within each type of fermions:

$$a_{ij}^d \cong a^d, \, a_{ij}^u \cong a^u, \, a_{ij}^l \cong a^l, \, a_{ij}^v \cong a^v \quad (4)$$

The first assumption result in n-1 massless particles and one massive particle with $m = na^F \eta$ (F=u,d,l,v) for each type of the SM fermions.

Because there is only one Higgs doublet which gives Dirac masses to all four types of fermions (up quarks, down quarks, charged leptons and neutrinos), it seems natural to make the *second assumption* [2, 3], namely, Yukawa constants for different types of fermions should be nearly equal:

$$a^d \approx a^u \approx a^l \approx a^v \approx a. \quad (5)$$

Taking into account the mass values for the third generation, the second assumption leads to the statement that *according to the flavor democracy the fourth SM family should exist*. In terms of the mass matrix above arguments mean

$$M^0 = a\eta \begin{pmatrix} 1 & 1 & 1 & 1 \\ 1 & 1 & 1 & 1 \\ 1 & 1 & 1 & 1 \\ 1 & 1 & 1 & 1 \end{pmatrix} \Rightarrow M^m = 4a\eta \begin{pmatrix} 0 & 0 & 0 & 0 \\ 0 & 0 & 0 & 0 \\ 0 & 0 & 0 & 0 \\ 0 & 0 & 0 & 1 \end{pmatrix}. (6)$$

Now, let us make the *third assumption*, namely, $a$ is between $e = g_w \sin\theta_W$ and $g_w/\cos\theta_W$. Therefore, the fourth family fermions are almost degenerate, in good agreement with experimental value $\rho=0.9998\pm0.0008$ [9], and their common mass lies between 320 GeV and 730 GeV. The last value is close to upper limit on heavy quark masses, $m_Q \leq 700$ GeV, which follows from partial-wave unitarity at high energies [10]. It is interest that with preferable value $a \approx g_w$ flavor democracy predicts $m_4 \approx 8 m_W \approx 640$ GeV. The masses of the first three family fermions, as well as an observable interfamily mixings, are generated due to the small deviations from the full flavor democracy [11, 12].

## 3. Arguments Against the Fifth SM Family

The first argument disfavoring the fifth SM family is the large value of $m_t \approx 175$ GeV. Indeed, partial-wave unitarity leads to $m_Q \leq 700$ GeV $\approx 4 m_t$ and in general we expect that $m_t << m_4 << m_5$. Then, neutrino counting at LEP results in fact that there are only three "light" ($2 m_\nu < m_Z$) non-sterile neutrinos, whereas in the case of five SM families four "light" neutrinos are expected.

The main restrictions on the new SM families come from experimental data on the parameters $\rho$ and $S$ (see [9] and references therein). The first one is sensitive to mass splitting of the up and down fermions, which is negligible according to the second assumption of the flavor democracy. The second one needs more detailed consideration. The contribution to $S$ from heavy degenerate SM family is equal to $2/3\pi$ ($\approx 0.21$), which should be compared with experimentally allowed value $\Delta S = -0.16 \pm 0.14$ [9]. At the first glance, the fourth and fifth SM families are excluded at 2.5 $\sigma$ and 4.1 $\sigma$, respectively. However, both the negative central value and comparatively small errors are caused mainly by asymmetries and R ratios measured at the Z pole and a number of them differ essentially from the SM predictions. For example, $A_{LR}$ deviation is 2.4 $\sigma$, $A_b$ is 2.5 $\sigma$ below the Standard model prediction *etcetera* (for details, see [9] and references therein). It would be useful to reanalyze experimental data excluding the mentioned observables. The rough estimations show that in this case the fourth SM family is allowed at 2 $\sigma$ and the fifth SM family is excluded at 3.5 $\sigma$ level. An enlargement of Higgs sector and/or the inclusion of Majorana mass terms for right-handed neutrinos may further improve the situation, but this is beyond the scope of present letter. Finally, the recent paper [13] which show that precision electroweak data allows the existence of a few extra families, if one allows neutral leptons to have masses close to 50 GeV, may be considered as an indication of the fact that the situation on the subject is far from clearness.

## 4. A Search for the Fourth SM Family

### 4.1. LHC

The fourth SM family quarks will be copiously produced at the LHC via gluon-gluon fusion (see [6] and references therein). The expected cross section is about 10(0.25) pb for a quark mass of 400(800) GeV. The fourth generation up-type quark, $u_4$, would predominantly decay via $u_4 \rightarrow Wb$, therefore, the expected event topologies are similar to those for $t$-quark pair production. The best channel for observing will be [14]:

$$u_4 \bar{u}_4 \to WWb\bar{b} \to (l\nu)(jj)b\bar{b},$$

where one $W$ decays leptonically and the other hadronically. The mass resolution is estimated to be 20(40) GeV for $m_4 = 320(640)$ GeV. The situation is much more complicate for down-type quark because the dominant decay mode is $d_4 \to Wt$ and the final state contains four $W$ bosons:

$$d_4 \bar{d}_4 \to tW^- \bar{t}W^+ \to bW^+W^- \bar{b}W^-W^+.$$

The small inter-family mixings [12] leads to the formation of the fourth family quarkonia. The most promising candidate for LHC is the pseudo-scalar quarkonium state, $\eta_4$, which will be produced resonantly via gluon-gluon fusion. Especially the decay channel $\eta_4 \to ZH$ is the matter of interest [15].

### 4.2. Future Lepton and Photon Colliders

The future lepton colliders will give opportunity to look for the fourth family leptons, which will hardly seen at hadron machines. Also, the number of different fourth family quarkonium states can be produced resonantly at lepton machines. Moreover, in difference from the LHC, states formed by up and down type quarks can be investigated separately even if their mass difference is small. The fourth family fermions, except of $\nu_4$, and various quarkonium states will be copiously produced at photon colliders.

### 4.3. Upgraded Tevatron

If the mass of the Higgs boson does not essentially exceed 200 GeV, the first indication of the fourth family may come soon from the Tevatron [7]. Indeed, the cross section of the Higgs boson production via gluon fusion is essentially enhanced due to extra heavy quarks [16]. For 100 GeV $< m_H < 200$ GeV the fourth SM family quarks with 300 GeV $< m_4 < 700$ GeV lead to the enhancement factor $k \approx 8$ [8]. If the fifth SM family exist, this factor becomes $k \approx 22$. Therefore, the search for the Higgs boson at the upgraded Tevatron can simultaneously result in the first indication of the fourth SM family. For illustration let me consider the process

$$\bar{p}p \to gg \to H \to W^*W^* \to l\nu jj \text{ and } l\bar{\nu}\bar{l}\nu$$

which was analyzed in [17]. Using the Figure 3 from this paper one can estimate the integrated luminosity values needed to reach:
a) 3σ statistical significance for discovery of the Higgs boson in three family case,
b) 3σ statistical significance for manifestation of the fourth family,
c) 5σ statistical significance for exclusion of the fifth family,
at different values of the Higgs mass, which are presented in Table 1. Therefore, if $m_H = 165$ GeV, the existence of the fifth SM family can be excluded at 5σ level by the recent Tevatron data.

In my opinion, the subject is sufficiently important in order to initiate the detailed studies, including different decay modes and detector aspects. For example, the search for H→ $\tau^+\tau^-$ channel will give an indication of the fourth family [7] and/or exclude the fifth family if 105 GeV < $m_H$ < 135 GeV.

Table 1.

| $m_H$, GeV | 120 | 135 | 150 | 165 | 180 | 195 |
|---|---|---|---|---|---|---|
| a) $L_{int}$, fb$^{-1}$ | 100 | 30 | 12 | 10 | 30 | 90 |
| b) $L_{int}$, fb$^{-1}$ | 12 | 4 | 1.5 | 1.2 | 4 | 11 |
| c) $L_{int}$, fb$^{-1}$ | 14 | 3.5 | 1.5 | 1 | 3.5 | 11 |

## 5. Alternatives

### 5.1. Two and More Higgs Doublets

In the framework of the Flavor Democracy hypothesis if there are only three SM families the large value of the t-quark mass to b-quark mass ratio ($m_t/m_b \approx 40$) may be natural if the masses of the up- and down-type quarks are generated by different Higgs doublets, as it takes place in the MSSM. In this case one can expect the following relation:

$$\tan\beta = \frac{v_u}{v_d} \approx \frac{m_t}{m_b},$$

where $v_u$ and $v_d$ are vacuum expectation values of the corresponding Higgs fields. Unfortunately, the MSSM contains huge number of free parameters, namely more than 160 for three MSSM families, and for this reason it seems more natural that SUSY should be realized at more fundamental, preonic or even pre-preonic level (for details see [18] and references therein).

Turning back to the SM with extended Higgs sector, let me finish this subsection with two remarks:

a) going further in this direction, one can assume that the masses of the charged leptons and neutrinos are generated by their own Higgs doublets,
b) introducing the isotriplet and vector isotriplet Higgs fields in addition to Higgs isodublet, one can change the tree level prediction ρ=1 and, therefore, relatively large mass splittings of the fourth family fermions may be allowed.

### 5.2. Exotic New Fermions

Another way to explain the relation $m_{b,\tau} << m_t$ is the introduction of exotic fermions. Let me consider as an example the extension of the SM fermion sector which is inspired by $E_6$ GUT model initially suggested by F. Gursey and collaborators [19]. It is known that this model is strongly favored in the framework of SUGRA (see [20] and references therein). For illustration let me restrict myself by quark sector:

$$\begin{pmatrix} u_L^0 \\ d_L^0 \end{pmatrix}, u_R^0, d_R^0; \quad \begin{pmatrix} c_L^0 \\ s_L^0 \end{pmatrix}, c_R^0, d_R^0; \quad \begin{pmatrix} t_L^0 \\ b_{Ll}^0 \end{pmatrix}, t_R^0, b_R^0$$

$$D_{1L}^0, D_{1R}^0; \quad D_{2L}^0, D_{2R}^0; \quad D_{3L}^0, D_{3R}^0.$$

According to Flavor Democracy the down quarks' mass matrix has the form:

$$M^0 = \begin{pmatrix} a\eta & a\eta & a\eta & a\eta & a\eta & a\eta \\ a\eta & a\eta & a\eta & a\eta & a\eta & a\eta \\ a\eta & a\eta & a\eta & a\eta & a\eta & a\eta \\ M & M & M & M & M & M \\ M & M & M & M & M & M \\ M & M & M & M & M & M \end{pmatrix},$$

where $M$ is the scale of "new" physics which determines the masses of the isosinglet quarks. As the result we obtain 5 massless quarks and the sixth quark has the mass $3M+m_t$.

## 6. Conclusion

There are two different approaches concerning the fourth SM family. The first one is following [9]: "Allowing arbitrary S, an extra generation of ordinary fermions is now excluded at the 99.2% CL. This is in agreement with a fit to the number of light neutrinos, $N_\nu=2.993\pm0.011$".

However, I prefer the moderate one [21]: "Today we have not any experimental indication of the new families. Precision electroweak data allow the one additional family at $2\sigma$ level. On the other hand, there are some arguments, including Flavor Democracy, favoring the fourth SM family. Therefore, let us wait the results from the LHC and at the same time carefully analyze the data on the Higgs boson search at the Tevatron".

## Acknowledgements

I am grateful to A. Celikel, A.K. Ciftci and I.F. Ginzburg for useful discussions and valuable remarks. I would like also to express my gratitude to DESY Directorate for invitation and hospitality.

## References

1. H. Harari, H. Haut and J. Weyers, Phys. Lett. B78(1978)459;
H. Fritzch, Nucl. Phys. B155(1979)189; B184(1987)391;
P. Kaus and S. Meshkov, Mod. Phys. Lett. A3(1988)1251;
H. Fritzch and J. Plankl, Phys. Lett. B237(1990)451.
2. A. Datta, Pramana 40(1993)L503.
3. A. Celikel, A.K. Ciftci and S. Sultansoy, Phys. Lett. B342(1995)257.
4. C.T. Hill and E.A. Paschos, Phys. Lett. B241(1990)96.
5. H. Fritzch, preprint MPI-Ph/92-42(1992), unpublished.
6. ATLAS Detector and Physics Performance Technical Design Report, CERN/LHCC/99-15(1999), p.663.
7. I.F. Ginzburg, I.P. Ivanov and A. Sciller, Phys. Rev. D 60(1999)095001.
8. E. Arik et al., ATLAS Internal Note ATL-PHYS-98-125(1999).

9. J. Erler and P. Langacker, in Review of Particle Physics, Eur. Phys. J. 3 (1998) 90.
10. M.S. Chanowitz, M.A. Furlan and I. Hinchliffe, Nucl. Phys. B153 (1979) 402.
11. A. Datta and S. Rayachaudhiri, Phys. Rev. D49 (1994) 4762.
12. S. Atag et al., Phys. Rev. D54 (1996) 5745.
13. M. Maltoni et al., Phys. Lett. B476 (2000) 107.
14. E. Arik et al., Phys. Rev. D58 (1998) 117701.
15. E. Arik et al., ATLAS Internal Note ATL-PHYS-99-061 (1999).
16. V.D. Barger and R.J.N. Philips, Collider Physics, Addison-Wisley, 1997.
17. T. Han and R.-J. Zhang, Phys. Rev. Lett. 82 (1999) 25.
18. S. Sultansoy, Ankara University preprint AU-HEP-00-01; hep-ph/0003269.
19. F. Gursey, P. Ramond and P. Sikivie, Phys. Lett. B60-(1976) 177; F. Gursey and M. Serdaroglu, Lett. Nuovo Cim. 21 (1978) 28.
20. J.L. Hewett and T.G. Rizzo, Phys. Reports 193 (1989) 193.
21. This paper.